\documentclass[notitlepage,nofootinbib,superscriptaddress,showpacs,preprintnumbers,amsmath,amssymb,prd,twocolumn,12 point]{revtex4-1}
\usepackage{epsfig,subfigure}
\usepackage{graphicx}
\usepackage{bbm}
\usepackage{enumerate}
\usepackage[colorlinks=true,citecolor=blue]{hyperref}
\draft 
\begin{document}
\author{Radha Raman Gautam}%
 \email{gautamrrg@gmail.com}
  \affiliation{Department of Physics, Himachal Pradesh University, Shimla 171005, INDIA.}%
\title{Trimaximal mixing with a texture zero}

\begin{abstract}
We analyze neutrino mass matrices having one texture zero, assuming that the neutrino mixing matrix has either its first (TM$_1$) or second (TM$_2$) column identical to that of the tribimaximal mixing matrix. We found that all the six possible one texture zero neutrino mass matrices are compatible with the present neutrino oscillation data when combined with TM$_1$ or TM$_2$ mixing. These textures have interesting predictions for the presently unknown parameters such as the effective Majorana neutrino mass, the Dirac CP violating phase and the neutrino mass scale. We also present a way to theoretically realize some of these textures using $A_4$ symmetry within the framework of type-I+II seesaw mechanism.
\end{abstract}
\keywords{Beyond Standard Model, Neutrino Physics, Discrete and Finite Symmetries}
\pacs{14.60.Pq, 11.30.Hv, 14.60.St}
\maketitle
\section{Introduction}
Various neutrino oscillation experiments in the last decade or so have measured the three lepton mixing angles and it is now clear that the flavor mixing in the lepton sector is quite large as compared to the quark sector. Non-Abelian discrete flavor symmetries have been extensively utilized to explain the large mixing in the lepton sector \cite{discrete}. Among the most widely studied lepton mixing patterns obtained using discrete non-Abelian symmetries is the tribimaximal (TBM) mixing pattern \cite{tbm}
\begin{equation}
U_{TBM} = \left(
\begin{array}{ccc}
-\frac{\sqrt{2}}{\sqrt{3}} & \frac{1}{\sqrt{3}} & 0 \\ 
\frac{1}{\sqrt{6}} & \frac{1}{\sqrt{3}} & -\frac{1}{\sqrt{2}}\\ 
\frac{1}{\sqrt{6}} & \frac{1}{\sqrt{3}} & \frac{1}{\sqrt{2}} 
\end{array}
\right)
\end{equation} 
which predicts the reactor mixing angle $\theta_{13}$ to be zero, a maximal atmospheric mixing angle, i.e., $\theta_{23} = \frac{\pi}{4}$, and the solar mixing angle $\theta_{12} = \sin^{-1}(1/\sqrt{3})$. 
However, recent neutrino oscillation experiments have found $\theta_{13}$ to be non-zero \cite{th13}, necessitating some modifications to the TBM mixing scheme to make it compatible with the present experimental data. In this context it has been observed that the first or the second column of TBM mixing matrix can still accommodate the recent neutrino oscillation data \cite{He}. When the second column of TBM mixing matrix remains intact while the other columns deviate from TBM values we denote that mixing matrix as TM$_2$ \cite{WR}. Similarly, if the first column of TBM remains intact and other columns deviate from TBM we denote it by TM$_1$ \cite{WR}. When either of these columns of TBM remain intact in the lepton mixing matrix we can parametrize the mixing matrix in terms of one mixing angle and one CP violating Dirac phase along with two Majorana phases \cite{He,WR}. \\
There are many other approaches which have been used to explain neutrino mixing, some of these are: texture zeros \cite{tz}, vanishing cofactors \cite{cofactor}, hybrid textures \cite{hybrid} etc. Out of these, texture zeros have been quite successful in explaining mixing in both the quark and the lepton sectors. In the basis where the charged lepton mass matrix is diagonal there are seven patterns of two texture zeros which are compatible with the current neutrino oscillation data \cite{tz}. There are six possible patterns of one texture zero in the neutrino mass matrix which are shown in Table \ref{tab1}. All of these are compatible with the present experimental data \cite{onezero}. \\
Neutrino mass matrices with two texture zeros in combination with TM$_2$ mixing have been studied in Ref. \cite{ourtm2} where it has been found that only two patterns namely $A_1$ and $A_2$ (with texture zeros at (1,1), (1,2) and (1,1), (1,3) entries, respectively) can satisfy the present experimental data when we combine two texture zeros with TM$_2$ mixing. The combination of TM$_1$ mixing with two texture zeros has been studied in Ref. \cite{ourtm1} and similar to the TM$_2$ case it has been found that only patterns $A_1$ and $A_2$ of two texture zeros are compatible with the present neutrino oscillation data when combined with TM$_1$ mixing. This approach of combining texture zeros with TM$_1$/TM$_2$ mixing turns out to be very fruitful as it leads to very predictive texture structures of neutrino mass matrix. In the present work we study neutrino mass matrices having one texture zero in combination with TM$_2$/TM$_1$ mixing.

\begin{table}[tb]
\begin{center}
\begin{tabular}{cc}
\hline
\hline
 Pattern  &        Constraining Equation         \\
 \hline
I &  $M_{ee}=0$    \\
II &  $M_{e \mu}=0$  \\
III &  $M_{e\tau}=0$    \\
IV &  $M_{\mu \mu}=0$  \\
V &  $M_{\mu\tau}=0$   \\
VI &  $M_{\tau\tau}=0$  \\
 \hline
 \hline
\end{tabular}
\end{center}
\caption{Six allowed patterns of one texture zero in the neutrino mass matrix.}
\label{tab1}
\end{table}

\section{TM$_2$ Mixing and one texture zero}
\subsection{TM$_2$ Mixing}
A neutrino mass matrix with TBM mixing can be diagonalized as 
\begin{equation}
M_{\text{diag}}=U_{TBM}^TM_{TBM}U_{TBM}
\end{equation}
where $U_{TBM}$ contains the eigenvectors $t=(-\frac{\sqrt{2}}{\sqrt{3}}~\frac{1}{\sqrt{6}}~\frac{1}{\sqrt{6}})^T$, $u=(\frac{1}{\sqrt{3}}~\frac{1}{\sqrt{3}}~\frac{1}{\sqrt{3}})^T$ and $v=(0~\frac{-1}{\sqrt{2}}~\frac{1}{\sqrt{2}})^T$. The diagonal matrix $M_{\text{diag}}$ is given as
\begin{equation}\label{eq:diag}
M_{\text{diag}} =\left(
\begin{array}{ccc}
 m_{1} & 0 & 0 \\
 0 & e^{2 i \alpha } m_{2} & 0 \\
 0 & 0 & e^{2 i \beta } m_{3}
\end{array}
\right),
\end{equation}
where $m_1$, $m_2$, and $m_3$ are the three neutrino masses and
$\alpha$ and $\beta$ are the two Majorana phases. The mass matrix $M_{TBM}$ is invariant under the transformations $G_t$, $G_u$ and $G_v$; i.e. $G_t^T M_{TBM} G_t=M_{TBM}$, $G_u^T M_{TBM} G_u=M_{TBM}$ and $G_v^T M_{TBM} G_v=M_{TBM}$ with $G_t=1-2tt^T$, $G_u=1-2uu^T$ and $G_v=1-2vv^T$. The transformation matrices $G_u$ and $G_v$ correspond to the magic symmetry \cite{Lam} and the $\mu-\tau$ symmetry \cite{Lam}, respectively. $G_i ~ (i=t,u,v)$ are generators of $Z_2$ group. A $3 \times 3$ neutrino mass matrix with TBM symmetry is invariant under the $Z^t_2 \times Z^u_2 \times Z^v_2$ group.
Recently, neutrino oscillation experiments have confirmed a non-zero $\theta_{13}$, thus, the neutrino mass matrix $M_{\nu}$ cannot remain invariant under the $\mu-\tau$ symmetry transformation $G_v$. However, the neutrino mass matrix can still be invariant under the magic symmetry transformation $G_u$ as the experimental data are still compatible with the magic symmetry. The mixing matrix which corresponds to the magic symmetry is known as the trimaximal mixing (TM$_2$) matrix and can be parametrized as \cite{He,WR,Bjorken,Lam,SK}
\begin{equation}\label{equtm2}
U_{\textrm{TM}_2}=
\left(
\begin{array}{ccc}
 \sqrt{\frac{2}{3}} \cos \theta &
   \frac{1}{\sqrt{3}} & \sqrt{\frac{2}{3}}
   \sin \theta \\
 -\frac{\cos\theta}{\sqrt{6}}+\frac{e^{-i \phi} \sin
\theta}{\sqrt{2}} & \frac{1}{\sqrt{3}} &
   -\frac{\sin\theta}{\sqrt{6}}-\frac{e^{-i \phi} \cos\theta}{\sqrt{2}} \\
 -\frac{\cos\theta}{\sqrt{6}}-\frac{e^{-i \phi}
   \sin \theta}{\sqrt{2}} &
   \frac{1}{\sqrt{3}} & -\frac{\sin
   \theta}{\sqrt{6}}
+\frac{e^{-i \phi}
   \cos \theta}{\sqrt{2}}\end{array}
\right).
\end{equation}
The TM$_2$ mixing matrix has its middle column fixed to the TBM value ($u$), which leaves only two free parameters ($\theta$ and $\phi$) in $U_{\text{TM}_2}$ after we take into account the unitarity constraints. The neutrino mass matrix corresponding to TM$_2$ mixing is given as
\begin{equation}\label{eq:recon}
M_{\textrm{TM}_2}=U_{\text{TM}_2}^*M_{\text{diag}}U_{\text{TM}_2}^{\dagger}.
\end{equation}

\subsection{One zero in $M_{\text{TM}_2}$}
A neutrino mass matrix with TM$_2$ mixing can be parameterized as \cite{magic}
\begin{equation}\label{eq:magic}
M_{\textrm{TM}_2} =\left(
\begin{array}{ccc}
 a & b & c \\
 b & d & a+c-d \\
 c & a+c-d & b-c+d
\end{array}
\right).
\end{equation}
The constraint equations for all the patterns with one texture zero and TM$_2$ mixing can be obtained
by substituting the respective texture zero constraints from Table \ref{tab1} into Eq. (\ref{eq:magic}).
In the diagonal charged lepton mass matrix basis, all the six patterns of one texture zero in the neutrino mass matrix are compatible with the present experimental data. The combination of these one texture zero patterns with TM$_2$ mixing is bound to produce very predictive forms of neutrino mass matrices.\\
The neutrino mass matrices with one texture zero and TM$_2$ mixing are given below:
\begin{equation}\label{eq:a1}
 M^{\textrm{I}}_{\textrm{TM}_2}= \left(
\begin{array}{ccc}
0 & b & c \\ b & d & c-d\\ c & c-d & b-c+d 
\end{array}
\right)
\end{equation}
\begin{equation}\label{eq:a2}
 M^{\textrm{II}}_{\textrm{TM}_2}= \left(
\begin{array}{ccc}
a & 0 & c \\ 0 & d & a+c-d\\ c & a+c-d & -c+d 
\end{array}
\right)
\end{equation}
\begin{equation}\label{eq:a3}
 M^{\textrm{III}}_{\textrm{TM}_2}= \left(
\begin{array}{ccc}
a & b & 0 \\ b & d & a-d\\ 0 & a-d & b+d 
\end{array}
\right)
\end{equation}
\begin{equation}\label{eq:a4}
 M^{\textrm{IV}}_{\textrm{TM}_2}= \left(
\begin{array}{ccc}
a & b & c \\ b & 0 & a+c\\ c & a+c & b-c 
\end{array}
\right)
\end{equation}
\begin{equation}\label{eq:a5}
 M^{\textrm{V}}_{\textrm{TM}_2}= \left(
\begin{array}{ccc}
a & b & c \\ b & a+c & 0\\ c & 0 & a+b 
\end{array}
\right)
\end{equation}
\begin{equation}\label{eq:a6}
 M^{\textrm{VI}}_{\textrm{TM}_2}= \left(
\begin{array}{ccc}
a & b & c\\ b & c-b & a+b \\ c & a+b & 0 
\end{array}
\right)
\end{equation}
The above mass matrices [Eq. (\ref{eq:a1}) to Eq. (\ref{eq:a6})] can be rewritten as:
\begin{equation}\label{eq:t1}
 M^{\textrm{I}}_{\textrm{TM}_2}= \left(
\begin{array}{ccc}
0 & b & c \\ b & c-\Delta & \Delta \\ c & \Delta & b-\Delta 
\end{array}
\right) \ \ \ \textrm{where} \ \ \ \Delta = c-d
\end{equation}
\begin{equation}\label{eq:t2}
 M^{\textrm{II}}_{\textrm{TM}_2}= \left(
\begin{array}{ccc}
a & 0 & c \\ 0 & c-\Delta & a+\Delta \\ c & a+\Delta & -\Delta 
\end{array}
\right) \ \ \ \textrm{where} \ \ \ \Delta = c-d
\end{equation}
\begin{equation}\label{eq:t3}
 M^{\textrm{III}}_{\textrm{TM}_2}= \left(
\begin{array}{ccc}
a & b & 0 \\ b & d & a-d\\ 0 & a-d & b+d 
\end{array}
\right)
\end{equation}
\begin{equation}\label{eq:t4}
 M^{\textrm{IV}}_{\textrm{TM}_2}= \left(
\begin{array}{ccc}
\Lambda - c & b & c \\ b & 0 & \Lambda\\ c & \Lambda & b-c 
\end{array}
\right)  \ \ \ \textrm{where} \ \ \ \Lambda = a+c
\end{equation}
\begin{equation}\label{eq:t5}
 M^{\textrm{V}}_{\textrm{TM}_2}= \left(
\begin{array}{ccc}
a & \Omega-a & c \\ \Omega-a & a+c & 0\\ c & 0 & \Omega 
\end{array}
\right) \ \ \ \textrm{where} \ \ \ \Omega = a+b
\end{equation}
\begin{equation}\label{eq:t6}
 M^{\textrm{VI}}_{\textrm{TM}_2}= \left(
\begin{array}{ccc}
\Omega -b & b & c\\ b & c-b & \Omega \\ c & \Omega & 0 
\end{array}
\right) \ \ \ \textrm{where} \ \ \ \Omega = a+b
\end{equation}
All the six patterns from Eq. (\ref{eq:t1}) to Eq. (\ref{eq:t6}) can be written as a linear combination of following matrices
\begin{small}
\begin{align}\label{eq:b}
& p_{12}= \left(
\begin{array}{ccc}
0 & 1 & 0\\ 1 & 0 & 0 \\ 0 & 0 & 1 
\end{array}
\right), \ \ \ 
p_{13}= \left(
\begin{array}{ccc}
0 & 0 & 1\\ 0 & 1 & 0 \\ 1 & 0 & 0 
\end{array}
\right), \nonumber \\ 
&p_{23}= \left(
\begin{array}{ccc}
1 & 0 & 0\\ 0 & 0 & 1 \\ 0 & 1 & 0 
\end{array}
\right), \ \ \
 b_{12}= \left(
\begin{array}{ccc}
-1 & 1 & 0\\ 1 & -1 & 0 \\ 0 & 0 & 0 
\end{array}
\right), \nonumber \\ 
&b_{13}= \left(
\begin{array}{ccc}
-1 & 0 & 1\\ 0 & 0 & 0 \\ 1 & 0 & -1 
\end{array}
\right), \ \ \ 
b_{23}= \left(
\begin{array}{ccc}
0 & 0 & 0\\ 0 & -1 & 1 \\ 0 & 1 & -1 
\end{array}
\right),
\end{align} 
\end{small}
where the first three are the symmetric permutation matrices and the last three are in block diagonal form, e.g. $M^{\textrm{I}}_{\textrm{TM}_2}$ is obtained as a linear combination of $p_{12}$, $p_{13}$ and $b_{23}$:
\begin{equation}
 M^{\textrm{I}}_{\textrm{TM}_2} = \Delta \ b_{23} +b \ p_{12} +c \ p_{13}.
\end{equation}
Similarly, we can construct other patterns. This representation brings all the patterns on equal footing; i.e., all the one texture zero patterns with TM$_2$ mixing are made up of simple combinations of two symmetric permutation matrices and a block diagonal matrix. The above decomposition into permutation and block diagonal matrices also helps in the symmetry realization of these patterns.\\ 
A neutrino mass matrix with TM$_2$ mixing is diagonalized by the mixing matrix $U = U_{\textrm{TM}_2}$ given in Eq. (\ref{equtm2}).
\begin{equation}
U_{\textrm{TM}_2}^T M_{\textrm{TM}_2} U_{\textrm{TM}_2} = M_{\textrm{diag.}}.
\end{equation} 
We can calculate the neutrino mixing angles from a given mixing matrix $U$ by using the following relations:
\begin{equation}\label{eqmixing}
s_{12}^{2}=\frac{|U_{12}|^{2}}{1-|U_{13}|^{2}}, s_{23}^{2}=\frac{|U_{23}|^{2}}{1-|U_{13}|^{2}}, \textrm{ and } s_{13}^{2}=|U_{13}|^{2}.
\end{equation}
The mixing angles for TM$_2$ mixing in terms of parameters $\theta$ and $\phi$ are
\begin{align}\label{eqth12}
s_{12}^{2} =& \frac{1}{3-2 \sin^2\theta}, \ \ \ s_{13}^{2} = \frac{2}{3}\sin^2\theta. \nonumber \\
s_{23}^{2} =& \frac{1}{2} \left(1+\frac{\sqrt{3} \sin 2 \theta \cos\phi}{3-2 \sin^2\theta}\right).
\end{align}
The Dirac CP violating phase $\delta$ can be obtained from the Jarlskog rephasing invariant ($J_{CP}$) \cite{jcp}
\begin{equation}
J_{CP} = \textrm{Im}(U_{11}U_{12}^*U_{21}^*U_{22}).
\end{equation}
In the standard parametrization 
\begin{equation}\label{eqjcp}
J_{CP}=s_{12}s_{23}s_{13}c_{12}c_{23}c_{13}^2 \sin \delta.
\end{equation}
For the TM$_2$ mixing matrix
\begin{equation}\label{eqjtm2}
J_{CP}=\frac{1}{6\sqrt{3}}\sin 2 \theta \sin \phi.
\end{equation}
From Eqs. (\ref{eqjcp}) and (\ref{eqjtm2}), we obtain
\begin{equation}\label{eqdelta}
\tan \delta = \frac{\cos 2\theta + 2}{2 \cos 2\theta+1} \tan \phi.
\end{equation}
The effective Majorana mass term relevant for neutrinoless double beta decay is given by
\begin{equation}
|M_{ee}|= |m_1  U^{2}_{e1}+m_2 e^{2 i \alpha } U^{2}_{e2}+m_3 e^{2 i \beta } U^{2}_{e3}|.
\end{equation}
For TM$_2$ mixing, the above expression takes the following form
\begin{equation}
|M_{ee}|= \frac{1}{3}|2 m_1 \cos ^2\theta + m_2 e^{2 i \alpha } + 2   m_3 e^{2 i \beta } \sin ^2 \theta |.
\end{equation}
There are many ongoing and forthcoming experiments such as GERDA \cite{gerda}, CUORE \cite{coure}, EXO \cite{exo}, NEXT \cite{next}, MAJORANA \cite{majorana}, SuperNEMO \cite{supernemo} which aim to achieve a sensitivity up to 0.01 eV for $|M_{ee}|$. Cosmological observations put an upper bound on the sum of neutrino masses 
\begin{equation}
\Sigma = \sum_{i=1,2,3}^{3} m_{i} .
\end{equation} 
Planck satellite data \cite{planck} combined with WMAP, CMB and BAO experiments limit the sum of neutrino masses $\sum m_{i}\leq 0.23$ eV at 95$\%$ confidence level (CL). In the present work, we assume a more conservative limit of $\sum m_{i}\leq 1$ eV.
The existence of one texture zero in the neutrino mass matrix with TM$_2$ mixing implies
\begin{equation}
(M_{\textrm{TM}_2})_{ij} = 0.
\end{equation}
This condition yields a complex equation viz.
\begin{equation}
m_1 A+m_2 B+m_3 C = 0
\end{equation}
where, $A = U_{a1}U_{b1}$, $B = U_{a2}U_{b2} e^{2 i \alpha}$, $C = U_{a3}U_{b3} e^{2 i \beta}$ and $a$, $b$ can take values $e$, $\mu$ and $\tau$.
The above complex equation yields two mass ratios:
\begin{equation}\label{eqm12}
\frac{m_1}{m_2} = \frac{\textrm{Re}(C) \textrm{Im}(B)-\textrm{Re}(B) \textrm{Im}(C)}{\textrm{Re}(A) \textrm{Im}(C)-\textrm{Re}(C) \textrm{Im}(A)}
\end{equation}
and
\begin{equation}\label{eqm13}
\frac{m_1}{m_3} = \frac{\textrm{Re}(C) \textrm{Im}(B)-\textrm{Re}(B) \textrm{Im}(C)}{\textrm{Re}(B) \textrm{Im}(A)-\textrm{Re}(A) \textrm{Im}(B)}
\end{equation}
where Re (Im) denotes the real (imaginary) part.
These mass ratios can be used to obtain the expression for the parameter $R_\nu$, which is the ratio of mass squared differences ($\Delta m_{ij}^2 = m_i^2 - m_j^2$):
\begin{equation}\label{eq:rnu}
R_\nu \equiv \frac{\Delta m_{21}^2}{|\Delta m_{31}^2|} = \frac{(\frac{m_2}{m_1})^2-1}{|(\frac{m_3}{m_1})^2-1|}
\end{equation}
where $m_1 > m_3$ for an inverted mass ordering (IO) and $m_1 < m_3$ for the normal mass ordering (NO).
For a texture zero to be compatible with the present neutrino oscillation data, the parameter $R_\nu$ should lie within its experimentally allowed range. The phenomenological predictions of patterns $M^{\textrm{II}}_{\textrm{TM}_2}$ and $M^{\textrm{III}}_{\textrm{TM}_2}$ are related and one can obtain the predictions for pattern $M^{\textrm{III}}_{\textrm{TM}_2}$ by making the following transformations:
\begin{equation}\label{eq:trans}
\theta_{23} \rightarrow \frac{\pi}{2}-\theta_{23}, \ \ \  \delta=\pi-\delta
\end{equation}
on the predictions of pattern $M^{\textrm{II}}_{\textrm{TM}_2}$ and vice-versa. This is because patterns $M^{\textrm{III}}_{\textrm{TM}_2}$ and $M^{\textrm{II}}_{\textrm{TM}_2}$ are related via 2-3 symmetry: $M^{\textrm{III}}_{\textrm{TM}_2} = p_{23}^T M^{\textrm{II}}_{\textrm{TM}_2} p_{23}$ where $p_{23}$ is the 2-3 permutation matrix given in Eq. (\ref{eq:b}). Similarly, patterns $M^{\textrm{IV}}_{\textrm{TM}_2}$ and $M^{\textrm{VI}}_{\textrm{TM}_2}$ are related to each other by above transformations. Thus, we need to study in detail only one of the 2-3 symmetry related patterns. \\
In the numerical analysis, the neutrino mass matrix is reconstructed using Eq. (\ref{eq:recon}), which takes into account the constraint of TM$_2$ mixing. For the numerical analysis we generate $10^8$ points ($10^{10}$ for pattern $M^{\textrm{V}}_{\textrm{TM}_2}$ with NO). The mass squared differences $\Delta m_{21}^2$ and $|\Delta m_{31}^2|$ are varied randomly within their 3$\sigma$ experimental ranges. Parameters $\theta$, $\phi$, $\alpha$ and $\beta$ are varied randomly within their full possible ranges. The texture zero constraint is imposed by requiring that the parameter $R_\nu$ in Eq. (\ref{eq:rnu}), written in terms of mass ratios $\frac{m_1}{m_2}$ and $\frac{m_1}{m_3}$ should lie within its 3$\sigma$ experimental range. In addition to above constraints, we require the allowed points to lie within the 3$\sigma$ experimental ranges of mixing angles $\theta_{12}$, $\theta_{13}$ and $\theta_{23}$ where the neutrino mixing angles are extracted by using the relations given in Eq. (\ref{eqmixing}). The experimental ranges of various neutrino oscillation parameters with their 1, 2, 3$\sigma$ ranges are given in Table \ref{tab2}. 
\begin{table*}[!]
\begin{center}
\begin{tabular}{|c|c|c|}
 \hline
Parameter & Normal Ordering & Inverted Ordering \\
  & best fit $\pm 1 \sigma,\pm 2 \sigma$ ~~ $3\sigma$ range  & best fit $\pm 1 \sigma,\pm 2 \sigma$~~~~~~ $3\sigma$ range \\
 \hline 
$\theta_{12}^{\circ}$ & $34.5^{+1.1,+2.3}_{-1.0,-2.0}$ ~~ $31.5$ - $38.0$ & $34.5^{+1.1,+2.3}_{-1.0,-2.0}$ ~~~~~~ $31.5$ - $38.0$ \\
$\theta_{23}^{\circ}$ & $41.0^{+1.1,+2.7}_{-1.1,-1.9}$  ~~$38.3$ - $52.8$ & $50.5^{+1.0,+1.7}_{-1.0,-2.3}$ ~~~~~~ $38.5$ - $53.0$ \\
$\theta_{13}^{\circ}$ & $8.44^{+0.18,+0.26}_{-0.15,-0.34}$  ~~$7.9$ - $8.9$ & $8.41^{+0.16,+0.29}_{-0.17,-0.41}$ ~~~~~~ $7.9$ - $8.93$ \\
$\delta_{CP}^{\circ}$ & $252^{+56,+99}_{-36,-99}$ ~~ $0.0$ - $360$ & ~~~~~~~ $259^{+47,+88}_{-41,-77}$ ~~~~ $0$ - $31$ $\oplus$ $142$ - $360$ \\
$\Delta m^{2}_{21}/10^{-5} eV^2 $ & $7.56^{+0.19,+0.39}_{-0.19,-0.36}$ ~~$7.05$ - $8.14$ & $7.56^{+0.19,+0.39}_{-0.19,-0.36}$ ~~~~~~ $7.05$ - $8.14$ \\
$|\Delta m^{2}_{3l}|/10^{-3} eV^2 $ & $2.55^{+0.04,+0.08}_{-0.04,-0.08}$ ~~ $2.43$ - $2.67$ & $2.49^{+0.04,+0.08}_{-0.04,-0.08}$ ~~~~~~ $2.37$ - $2.61$ \\
 \hline 
 \end{tabular}
\caption{Current neutrino oscillation parameters from global fits \cite{data}.}
\label{tab2}
\end{center}
\end{table*}

\begin{figure*}[t]
\centering 
\includegraphics[scale=0.3]{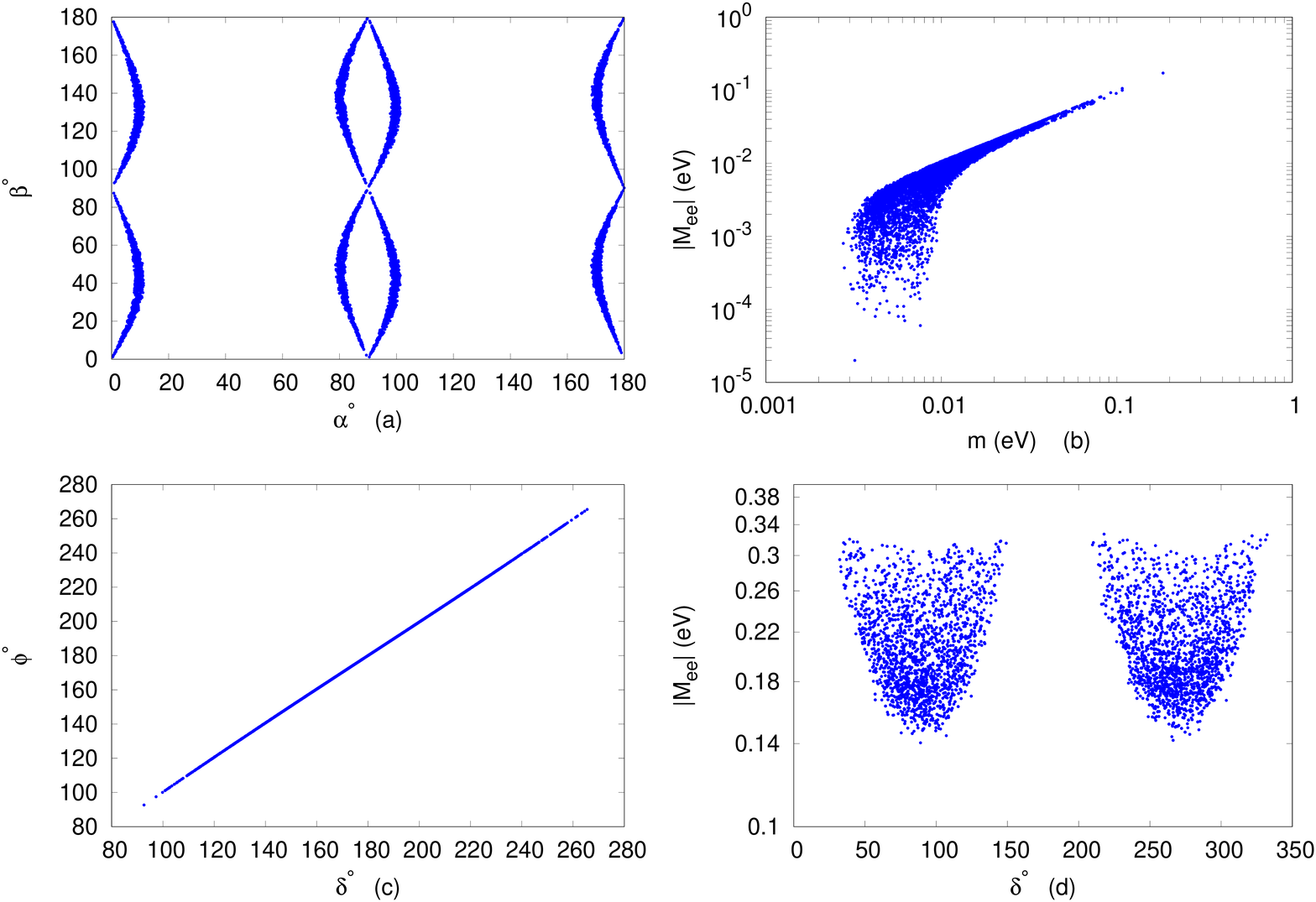}
\caption{Correlation plots for patterns $M^{\textrm{I}}_{\text{TM}_2}$ (a), $M^{\textrm{II}}_{\text{TM}_2}$ (b), $M^{\textrm{IV}}_{\text{TM}_2}$ (c) and $M^{\textrm{V}}_{\text{TM}_2}$ (d) with NO.}
\label{fig1}
\end{figure*} 
The numerical results for one texture zero in the neutrino mass matrix with TM$_2$ mixing are presented below. The main observations are:
\begin{enumerate}[i]
\item All six patterns of one texture zero in the neutrino mass matrix with TM$_2$ mixing are consistent with the present neutrino oscillation data.
\item The pattern $M^{\textrm{I}}_{\textrm{TM}_2}$ is consistent with normal mass ordering only.
\item All the viable patterns except $M^{\textrm{I}}_{\textrm{TM}_2}$, allow a quasidegenerate mass spectrum.
\item In case of NO, the parameter $|M_{ee}|$ can vanish for patterns $M^{\textrm{I}}_{\textrm{TM}_2}$, $M^{\textrm{II}}_{\textrm{TM}_2}$ and $M^{\textrm{III}}_{\textrm{TM}_2}$. For the remaining patterns $|M_{ee}|$ is found to be bounded from below.
\item The smallest neutrino mass cannot vanish except for patterns $M^{\textrm{II}}_{\textrm{TM}_2}$ and $M^{\textrm{III}}_{\textrm{TM}_2}$ with IO. 
\item The parameter $J_{CP}$ cannot vanish for patterns $M^{\textrm{II}}_{\text{TM}_2}$ with IO and $M^{\textrm{V}}_{\text{TM}_2}$ with NO, implying that these patterns are necessarily CP violating. 
\item The atmospheric neutrino mixing angle $\theta_{23}$ remains below (above) 45$^\circ$ for pattern $M^{\textrm{IV}}_{\textrm{TM}_2}$ ($M^{\textrm{VI}}_{\textrm{TM}_2}$) with NO.
\end{enumerate}
\begin{figure*}[t]
\centering 
\includegraphics[scale=0.3]{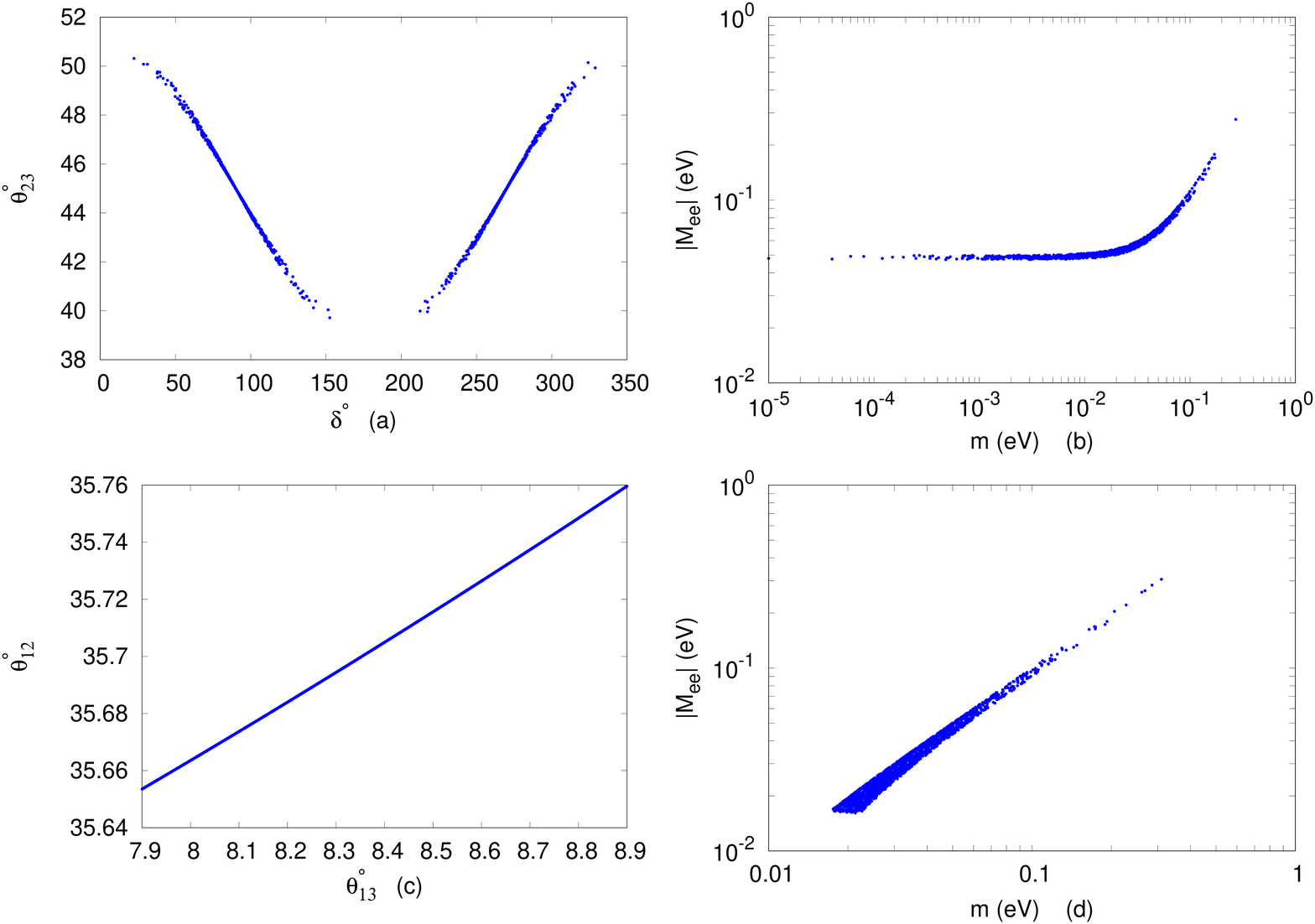}
\caption{Correlation plots for patterns $M^{\textrm{II}}_{\text{TM}_2}$ (a,b), $M^{\textrm{IV}}_{\text{TM}_2}$ (c) and $M^{\textrm{V}}_{\text{TM}_2}$ (d) with IO.}
\label{fig2}
\end{figure*} 
\begin{figure*}[t]
\centering 
\includegraphics[scale=0.3]{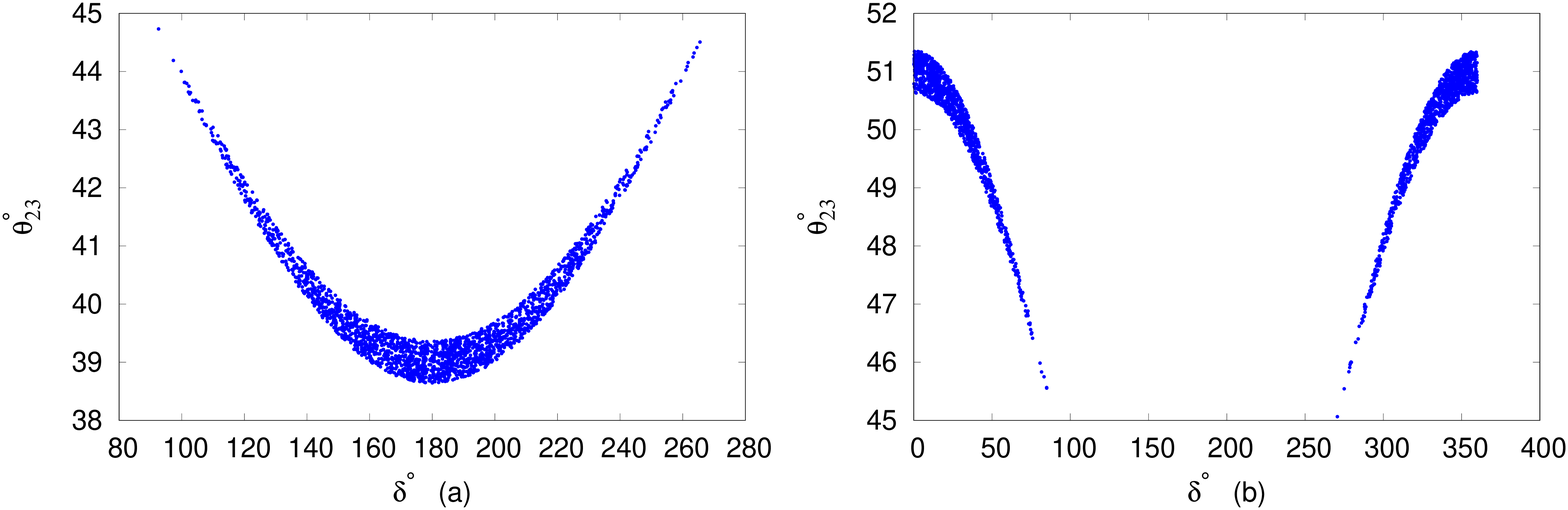}
\caption{Correlation plots for patterns $M^{\textrm{IV}}_{\text{TM}_2}$ (a) and $M^{\textrm{VI}}_{\text{TM}_2}$ (b) with NO.}
\label{fig3}
\end{figure*} 
The numerical predictions for the presently unknown neutrino parameters have been summarized in Table \ref{tab3}. The allowed ranges (at 3$\sigma$ CL) of parameters $\theta_{12}$, $\theta$ and $J_{CP}$ are ($35.65^{\circ}$ - $35.76^{\circ}$), ($9.68^{\circ}$ - $10.93^{\circ}$) and ($-0.0363$ - $0.0363$), respectively, for all the allowed patterns except for patterns $M^{\textrm{II,\ III}}_{\text{TM}_2}$ with IO and $M^{\textrm{V}}_{\text{TM}_2}$ with NO, for which $J_{CP}$ cannot vanish and has the allowed ranges $\pm$($0.006$ - $0.036$) and $\pm$($0.0165$ - $0.036$), respectively. Some of the interesting correlation plots are given in Figs. \ref{fig1}, \ref{fig2} and \ref{fig3}. Fig. \ref{fig1}(a) shows that the Majorana phases $\alpha$ and $\beta$ are strongly correlated with each other for pattern $M^{\textrm{I}}_{\text{TM}_2}$ with NO. One can see from Fig. \ref{fig1}(c) that the Dirac phase $\delta$ and phase $\phi$ are linearly correlated and are almost identical to each other. From Eq. (\ref{eqdelta}) we can see that the ratio $(\frac{\cos 2\theta + 2}{2 \cos 2\theta + 1})$ multiplying with $\tan \phi$ remains $\approx 1$ for the allowed values of $\theta$. This leads to the feature $\delta \approx \phi$, for all the neutrino mass matrices with a texture zero and TM$_2$ mixing.\\
For pattern $M^{\textrm{II}}_{\text{TM}_2}$ the Dirac phase $\delta$ is restricted to two regions [Fig. \ref{fig2}(a)]. The correlation between mixing angles $\theta_{13}$ and $\theta_{12}$ is shown in Fig. \ref{fig2}(c). This is a generic feature of TM$_2$ mixing arising from Eq. (\ref{eqth12}). Since the TBM value of $\theta_{12}$ is already above its experimental best fit value, an increase in $\theta$ drives $\theta_{12}$ further away from the best fit experimental value. Thus, TM$_2$ mixing leads to some tension with mixing angle $\theta_{12}$. Fig. \ref{fig3} shows the 2-3 symmetry between patterns $M^{\textrm{IV}}_{\text{TM}_2}$ and $M^{\textrm{VI}}_{\text{TM}_2}$.
\begin{table*}[t]
\centering
\begin{tabular}{|c|c|c|c|c|c|}
 \hline 
Pattern     & Mass Ordering & $|M_{ee}|$ (eV)& Lightest Neutrino Mass $m$ (eV)   & $\sum m_i$ (eV) & $\delta^{\circ}$ \\
 \hline
I& NO       & $0.0$            & $0.0026$ - $0.0084$& $0.061$ - $0.073$ & $0$ - $360$         \\
             \hline 
II&NO      & $0.0$ - $0.172$  & $0.002$ - $0.183$  & $0.056$ - $0.57$  & $0$ - $360$  \\
            & IO       & $0.047$ - $0.275$& $0.0$ - $0.272$    & $0.097$ - $0.825$ & $21$ - $154$ $\oplus$ $200$ - $345$ \\
 \hline
III&NO     & $0.0$ - $0.172$  & $0.002$ - $0.183$  & $0.056$ - $0.57$  & $0$ - $360$  \\ 
            & IO       & $0.047$ - $0.275$& $0.0$ - $0.272$    & $0.097$ - $0.825$ & $29$ - $159$ $\oplus$ $200$ - $332$  \\
 \hline  
IV&NO   & $0.034$ - $0.237$& $0.035$ - $0.3$    & $0.13$ - $0.9$    & $90$ - $270$  \\
            & IO       & $0.015$ - $0.126$& $0.007$ - $0.18$   & $0.106$ - $0.553$ & $0$ - $360$  \\
 \hline
V& NO &      $0.14$ - $0.34$  &    $0.147$ - $0.332$       &   $0.45$ - $1$      &  $32$ - $150$ $\oplus$ $209$ - $333$    \\
            & IO       & $0.015$ - $0.305$& $0.017$ - $0.31$   & $0.127$ - $0.94$  & $0$ - $360$ \\
 \hline
VI&NO & $0.034$ - $0.237$& $0.035$ - $0.3$    & $0.13$ - $0.9$    & $0$ - $90$ $\oplus$ $270$ - $360$  \\ 
            & IO       & $0.015$ - $0.126$& $0.007$ - $0.18$   & $0.106$ - $0.553$ & $0$ - $360$ \\
 \hline
\end{tabular}
\caption{Numerical predictions (at 3$\sigma$ CL) for patterns having one texture zero in $M_\nu$ with TM$_2$ mixing.}\label{tab3}
\end{table*}
\subsection{Symmetry realization}
The neutrino mass matrices with one texture zero and TM$_2$ mixing can be realized in the framework of type-I+II seesaw mechanism \cite{seesaw1,seesaw2} using $A_4$ \cite{a4} symmetry. In addition to the three standard model lepton $SU(2)_L$ doublets $D_{l_L}$ (where $l=e,\mu~ \text{and}~ \tau$) and three right-handed charged lepton singlets $l_R$, we need a $SU(2)_L$ singlet right-handed neutrino $\nu_R$, six $SU(2)_L$ doublet Higgs fields $\psi_i$ and $\varphi_i$ ($i = 1, 2, 3$), and two $SU(2)_L$ triplet Higgs fields $\Delta_1$, $\Delta_2$. We also impose an additional $Z_2$ symmetry, to prevent couplings between charged leptons (neutrinos) and scalars $\varphi_i$ ($\psi_i$).
Below we discuss in detail the symmetry realization of pattern $M^{\textrm{I}}_{\text{TM}_2}$. The transformation properties of various fields under $A_4$ and $Z_2$ corresponding to pattern I are given in Table \ref{tab4}. These transformation properties lead to the following Yukawa Lagrangian which is invariant under $A_4$ and $Z_2$.
\begin{widetext}
\begin{eqnarray}
-\mathcal{L}_{\text{Yukawa}} & = & y_1 (\overline{D}_{e_L} \psi_1 + \overline{D}_{\mu_L} \psi_2 
+ \overline{D}_{\tau_L} \psi_3)_{\underline{1}} e_{R_{\underline{1}}}   +  y_2 (\overline{D}_{e_L} \psi_1 + \omega^2 
\overline{D}_{\mu_L} \psi_2 +  \omega \overline{D}_{\tau_L} \psi_3)_{\underline{1}^\prime} \tau_{R_{\underline{1}^{\prime\prime}}}  \nonumber \\
& + & y_3 (\overline{D}_{e_L} \psi_1 + \omega \overline{D}_{\mu_L} \psi_2 + \omega^2 \overline{D}_{\tau_L} \psi_3)_{\underline{1}^{\prime \prime}} \mu_{R_{\underline{1}^{\prime}}} +  y_4 (\overline{D}_{e_L} \tilde{\varphi}_1 + \overline{D}_{\mu_L} \tilde{\varphi}_2 + \overline{D}_{\tau_L} \tilde{\varphi}_3)_{\underline{1}} \nu_{R_{\underline{1}}}  \nonumber \\ 
& - & y_{\Delta_1} (D_{e_L}^T C^{-1} D_{e_L} + \omega D_{\mu_L}^T C^{-1} D_{\mu_L}   +  \omega^2 D_{\tau_L}^T C^{-1} D_{\tau_{L}})_{\underline{1}^{\prime \prime}} i \tau_2 \Delta_{1_{\underline{1}^{\prime}}} \nonumber \\ 
& - & y_{\Delta_2} (D_{e_L}^T C^{-1} D_{e_L} + \omega^2 D_{\mu_L}^T C^{-1} D_{\mu_L}  + \omega D_{\tau_L}^T C^{-1} D_{\tau_{L}})_{\underline{1}^{\prime}} i \tau_2 \Delta_{2_{\underline{1}^{\prime \prime}}} - m_R( \nu_{R}^T C^{-1} \nu_{R}) + \ \ \textrm{H.c.}
\end{eqnarray}
\end{widetext}
where $\tilde{\varphi}=i \tau_2 \varphi^*$. Assuming that $\psi_i$ Higgs fields acquire non-zero vacuum expectation values (VEVs) along the direction $\langle \psi \rangle_o = v_{\psi}(1, 1, 1)^T$, leads to the following form of the charged lepton mass matrix
\begin{equation}
 m_l = \left(
\begin{array}{ccc}
y_1 v_{\psi} & y_2 v_{\psi} & y_3 v_{\psi} \\ 
y_1 v_{\psi} & y_2 \omega v_{\psi} & y_3 \omega^2 v_{\psi} \\
y_1 v_{\psi} & y_2 \omega^2 v_{\psi} & y_3 \omega v_{\psi}
\end{array}
\right).
 \end{equation}   
The $\varphi_i$ fields which couple to neutrinos are assumed to have the VEV alignment: $\langle \varphi \rangle_o = v_{\varphi}(0, -1, 1)^T$. Similar vacuum alignment has been obtained earlier in Ref. \cite{vev} for $A_4$ and $SU(2)_L$ triplet scalars. The above VEV alignment leads to the following Dirac neutrino mass matrix.
\begin{equation}
 m_D = y_4 v_{\varphi} (0,-1,1)^T.
 \end{equation}
 With only one right-handed neutrino with mass $m_R$, the effective neutrino mass matrix obtained using the type-I seesaw mechanism $m^I_{\nu} \approx m_D m_R^{-1} m_D^T$, has the form
 \begin{equation}
 m^I_{\nu} = a \left(
\begin{array}{ccc}
0 & 0 & 0 \\ 0 & 1 & -1 \\ 0& -1 & 1
\end{array}
\right) \ \ \ \ \textrm{where} \ \  a=y_4^2 v_{\varphi}^2/m_R.
\end{equation}
The type-II seesaw contribution to the effective neutrino mass matrix has the following form which is obtained when the $SU(2)_L$ triplet Higgses $\Delta_1$ and $\Delta_2$ acquire non-zero and small VEVs:
\begin{equation}
 m^{II}_{\nu} = \left(
\begin{array}{ccc}
b+c & 0 & 0 \\ 0 & \omega b + \omega^2 c & 0 \\ 0& 0 & \omega^2 b + \omega c
\end{array}
\right)
\end{equation}
where $b=y_{\Delta_1} v_{\Delta_1}$ and $c=y_{\Delta_2} v_{\Delta_2}$.
The collective effective neutrino mass matrix $m_\nu = m_\nu^I + m_\nu^{II}$ from the type-I+II seesaw mechanism becomes
\begin{equation}
 m_\nu = \left(
\begin{array}{ccc}
b+c & 0 & 0 \\ 0 & a + \omega b + \omega^2 c & -a \\ 0 & -a & a + \omega^2 b + \omega c
\end{array}
\right).
\end{equation}
In the present basis the charged lepton mass matrix is non-diagonal. We move to the diagonal charged lepton mass matrix basis using the transformation $M_l = U_L^\dagger m_l U_R$, where 
\begin{equation}
 U_L=\frac{1}{3}\left(
\begin{array}{ccc}
1 & 1 & 1 \\ 1 & \omega & \omega^2 \\ 1 & \omega^2 & \omega
\end{array}
\right)
\end{equation}
and $U_R$ is a unit matrix.
In this basis the effective neutrino mass matrix becomes
\begin{equation}
 M_\nu = \left(
\begin{array}{ccc}
0 & b & c \\ 
b & -a+c & a \\ 
c & a & -a+b
\end{array}
\right)
\end{equation}
which is the patten $M^{\textrm{I}}_{\text{TM}_2}$ of one texture zero with TM$_2$ mixing. The symmetry realization of other patterns can be obtained in a similar way. We have summarized the desired transformation properties of various leptonic and scalar fields (under $A_4$ and $Z_2$) which lead to neutrino mass matrices with one texture zero and TM$_2$ mixing, in Table \ref{tab4}. \\
For the symmetry realization of above patterns, we require many Higgs $SU(2)_L$ doublets. It should be noted that such multi-Higgs models generally lead to flavor changing neutral currents which can contribute to charged lepton flavor violating decays. However, an explicit calculation of such effects is beyond the scope of the present work.
\begin{table*}[t]
\begin{center}
\begin{tabular}{|c|c|c|c|c|c|c|c|c|c|c|c|}
\hline 
Pattern& Symmetry & $D_{l_{L}}$ & $e_R$& $\mu_R$ & $\tau_R $ & $\nu_{R}$ & $\psi $ & $\varphi $ & $\Delta_1$&$\Delta_2$ &$D_{L}$ Triplet Representation under $A_4$ \\ 
\hline 
&$SU(2)_L$ & 2 & 1 &1&1& 1 & 2 & 2 & 2&3 & \\ 
$M^{\textrm{I}}_{\textrm{TM}_2}$&$A_4$ & 3 & $1$ & $ 1^\prime$ & $1^{\prime \prime}$ & 1 & 3 & 3 & $1^{\prime}$ & $1^{\prime \prime}$& $3 \sim \left(\begin{array}{c}D_{e_{L}} \\ D_{\mu_{L}}  \\ D_{\tau_{L}} \end{array}\right)$  \\ 
&$Z_2$ & 1 & 1&1&1 & -1 & 1 & -1 & 1& 1&\\ 
\hline 
&$SU(2)_L$ & 2 & 1 &1&1& 1 & 2 & 2 & 3& 3 & \\ 
$M^{\textrm{II}}_{\textrm{TM}_2}$&$A_4$ & 3 & $1$& $1^\prime$&$1^{\prime \prime}$ & 1 & 3 & 3 & $1$ & $1^{\prime \prime}$ & $3 \sim \left(\begin{array}{c}D_{e_{L}} \\ D_{\mu_{L}}  \\ D_{\tau_{L}} \end{array}\right)$  \\ 
&$Z_2$ & 1 & 1 &1&1& -1 & 1 & -1 & 1& 1&\\ 
\hline 
&$SU(2)_L$ & 2 & 1&1&1 & 1 & 2 & 2 & 3& 3 & \\ 
$M^{\textrm{III}}_{\textrm{TM}_2}$&$A_4$ & 3 & $1$& $ 1^\prime$& $1^{\prime \prime}$ & 1 & 3 & 3 & $1$ & $1^{\prime}$&  $3 \sim \left(\begin{array}{c}D_{e_{L}} \\ D_{\mu_{L}}  \\ D_{\tau_{L}} \end{array}\right)$ \\ 
&$Z_2$ & 1 & 1&1&1 & -1 & 1 & -1 & 1& 1&\\ 
\hline 
&$SU(2)_L$ & 2 & 1&1&1 & 1 & 2 & 2 & 3& 3 & \\ 
$M^{\textrm{IV}}_{\textrm{TM}_2}$&$A_4$ & 3 & $1^{\prime \prime}$& $ 1$& $ 1^\prime$ & 1 & 3 & 3 & $1^{\prime}$ &$1^{\prime \prime}$ & $3 \sim \left(\begin{array}{c}D_{\mu_{L}} \\ D_{\tau_{L}}  \\ D_{e_{L}} \end{array}\right)$\\ 
&$Z_2$ & 1 & 1 & 1&1&-1 & 1 & -1 & 1& 1&\\ 
\hline 
&$SU(2)_L$ & 2 & 1&1&1 & 1 & 2 & 2 & 3& 3 & \\ 
$M^{\textrm{V}}_{\textrm{TM}_2}$&$A_4$ & 3 & $1^\prime$& $1^{\prime \prime}$& $ 1$& 1 & 3 & 3 & 1 & $1^{\prime}$ & $3 \sim \left(\begin{array}{c}D_{\tau_{L}} \\ D_{e_{L}}  \\ D_{\mu_{L}} \end{array}\right)$\\ 
&$Z_2$ & 1 & 1 &1&1& -1 & 1 & -1 & 1& 1 &\\ 
\hline  
&$SU(2)_L$ & 2 & 1 &1&1& 1 & 2 & 2 & 3& 3 & \\ 
$M^{\textrm{VI}}_{\textrm{TM}_2}$&$A_4$ & 3 & $1^\prime$& $1^{\prime \prime}$& $ 1$ & 1 & 3 & 3 & $1^{\prime \prime}$ & $1^{\prime}$ & $3 \sim \left(\begin{array}{c}D_{\tau_{L}} \\ D_{e_{L}}  \\ D_{\mu_{L}} \end{array}\right)$\\ 
&$Z_2$ & 1 & 1 & 1&1&-1 & 1 & -1 & 1& 1&\\ 
\hline 
\end{tabular}
\end{center}
\caption{Transformation properties of various fields under $A_4$ and $Z_2$. The VEV alignments for $\psi$ and $\varphi$ are $\langle \psi \rangle_o = v_{\psi}(1, 1, 1)^T$ and $\langle \varphi \rangle_o = v_{\varphi}(0, -1, 1)^T$.
}
\label{tab4}
\end{table*} 
\section{TM$_1$ Mixing and one texture zero}
\subsection{TM$_1$ Mixing}
In section II we studied the one texture zero patterns having TM$_2$ mixing. For these patterns the allowed values of $\theta_{12}$ lie in the 2$\sigma$ upper limit and values within the 1$\sigma$ experimental range are not allowed. This leads to some tension with the present neutrino oscillation data. However, this is a generic feature of TM$_2$ mixing. One possible way to resolve this discrepancy is to consider charged lepton corrections to the neutrino mixing matrix. \\
Alternatively, instead of considering TM$_2$ mixing one may also consider TM$_1$ mixing where $\theta_{12}$ can take values which are in good agreement with the present neutrino oscillation data. In this section we explore the neutrino mass matrices having one texture zero along with TM$_1$ mixing. The TM$_1$ mixing matrix can be parametrized as \cite{He,WR,Lam,SK}:
\begin{equation}\label{eq:tm1}
U_{\text{TM}_1}=\left(
\begin{array}{ccc}
 \sqrt{ \frac{2}{3}} & \frac{1}{\sqrt{3}} \cos \theta &
\frac{1}{\sqrt{3}} \sin \theta \\
  -\frac{1}{\sqrt{6}} &
      \frac{1}{\sqrt{3}}\cos\theta-\frac{e^{i \phi} \sin
\theta}{\sqrt{2}} &
\frac{1}{\sqrt{3}}\sin\theta+
\frac{e^{i \phi} \cos\theta}{\sqrt{2}} \\
  -\frac{1}{\sqrt{6}} &
 \frac{1}{\sqrt{3}}\cos\theta+\frac{e^{i \phi}
                        \sin \theta}{\sqrt{2}} &
 \frac{1}{\sqrt{3}}\sin \theta
-\frac{e^{i \phi}
   \cos \theta}{\sqrt{2}}\end{array}
\right)
\end{equation}
here, the first column of the neutrino mixing matrix is identical with TBM mixing matrix and the other two columns have been parametrized in terms of two free parameters ($\theta$ and $\phi$) after taking into consideration the unitarity constraints on the mixing matrix. The corresponding neutrino mass matrix for TM$_1$ mixing is given as
\begin{equation}\label{eq:reco}
M_{\text{TM}_1}=U_{\text{TM}_1}^*M_{\text{diag}}U_{\text{TM}_1}^{\dagger}.
\end{equation}
\subsection{One zero in $M_{\text{TM}_1}$}
A neutrino mass matrix with TM$_1$ mixing can be written as
\begin{equation}\label{eq:mtm1}
M_{\textrm{TM}_1} =\left(
\begin{array}{ccc}
 a & 2 b &2 c \\
 2 b & 4b+d & a-b-c-d \\
 2 c & a-b-c-d & 4c+d
\end{array}
\right).
\end{equation}
All the neutrino mass matrices with one texture zero and TM$_1$ mixing can be obtained by substituting the respective constraints from Table \ref{tab1} in Eq. (\ref{eq:mtm1}):
\begin{equation}\label{eq:at1}
 M_{\text{TM}_1}^{\textrm{I}}= \left(
\begin{array}{ccc}
0 & 2 b & 2 c \\ 2 b &4 b+d & -b-c-d\\2 c & -b-c-d & 4 c+d 
\end{array}
\right)
\end{equation}
\begin{equation}\label{eq:at2}
 M_{\text{TM}_1}^{\textrm{II}}= \left(
\begin{array}{ccc}
a & 0 & 2 c \\ 0 & d & a-c-d\\ 2 c & a-c-d & 4c+d 
\end{array}
\right)
\end{equation}
\begin{equation}\label{eq:at3}
 M_{\text{TM}_1}^{\textrm{III}}= \left(
\begin{array}{ccc}
a & 2b & 0 \\ 2b & 4b+d & a-b-d\\ 0 & a-b-d & d 
\end{array}
\right)
\end{equation}
\begin{equation}\label{eq:at4}
 M_{\text{TM}_1}^{\textrm{IV}}= \left(
\begin{array}{ccc}
a & 2b & 2c \\ 2b & 0 & a+3b-c\\ 2c & a+3b-c & 4c-4b 
\end{array}
\right)
\end{equation}
\begin{equation}\label{eq:at5}
 M_{\text{TM}_1}^{\textrm{V}}= \left(
\begin{array}{ccc}
a & 2b & 2c \\ 2b & a+3b-c & 0\\ 2c & 0 & a+3c-b 
\end{array}
\right)
\end{equation}
\begin{equation}\label{eq:at6}
 M_{\text{TM}_1}^{\textrm{VI}}= \left(
\begin{array}{ccc}
a & 2b & 2c\\ 2b & 4b-4c & a+3c-b \\ 2c & a+3c-b & 0 
\end{array}
\right).
\end{equation}
A neutrino mass matrix with TM$_1$ mixing can be diagonalized by the mixing matrix $U = U_{\textrm{TM}_1}$ given in Eq. (\ref{eq:tm1}).
\begin{equation}
U_{\textrm{TM}_1}^T M_{\textrm{TM}_1} U_{\textrm{TM}_1} = M_{\textrm{diag.}}.
\end{equation} 
The mixing angles for TM$_1$ mixing in terms of parameters $\theta$ and $\phi$ are
\begin{align}\label{eqth122}
&s_{13}^{2}=\frac{1}{3}\sin^2\theta, \ \ \ s_{12}^{2} = 1-\frac{2}{3-\sin^2\theta},
\nonumber \\
&s_{23}^{2}=\frac{1}{2} \left(1+\frac{\sqrt{6}   \sin 2 \theta \cos\phi}{3-\sin^2\theta}\right).
\end{align}
We see from Eq. (\ref{eqth122}) that the solar mixing angle $\theta_{12}$ is smaller than its TBM value $s_{12}^{2}=1/3$. In contrast, for TM$_2$ mixing, the value of $\theta_{12}$ is larger than the TBM value. Since the experimental best fit value of $\theta_{12}$ is towards the lower side of the TBM value, TM$_1$ mixing is more appealing than TM$_2$ mixing.
The Dirac CP violating phase $\delta$ can be obtained from the Jarlskog rephasing invariant ($J_{CP}$) \cite{jcp}
\begin{equation}
J_{CP} = \textrm{Im}(U_{11}U_{12}^*U_{21}^*U_{22}).
\end{equation}
For the TM$_1$ mixing matrix
\begin{equation}\label{eqjtm1}
J_{CP}=\frac{1}{6\sqrt{6}}\sin 2 \theta \sin \phi.
\end{equation}
Using Eqs. (\ref{eqjcp}) and (\ref{eqjtm1}), we obtain
\begin{equation}\label{eqdelta2}
\tan \delta = \frac{\cos 2\theta+ 5}{5 \cos 2\theta + 1} \tan \phi.
\end{equation}
The effective Majorana mass for TM$_1$ mixing is given by
\begin{equation}
|M_{ee}|= \frac{1}{3} |2 m_1+ m_2 e^{2 i \alpha } \cos ^2 \theta +  m_3 e^{2 i \beta } \sin ^2 \theta |.
\end{equation}
The existence of one texture zero in the neutrino mass matrix implies
\begin{equation}
(M_{\textrm{TM}_1})_{ij} = 0.
\end{equation}
Following the same procedure as we did for TM$_2$ mixing, we analyse the phenomenological predictions of neutrino mass matrices having one texture zero and TM$_1$ mixing.\\
\begin{figure*}[t]
\centering 
\includegraphics[scale=0.3]{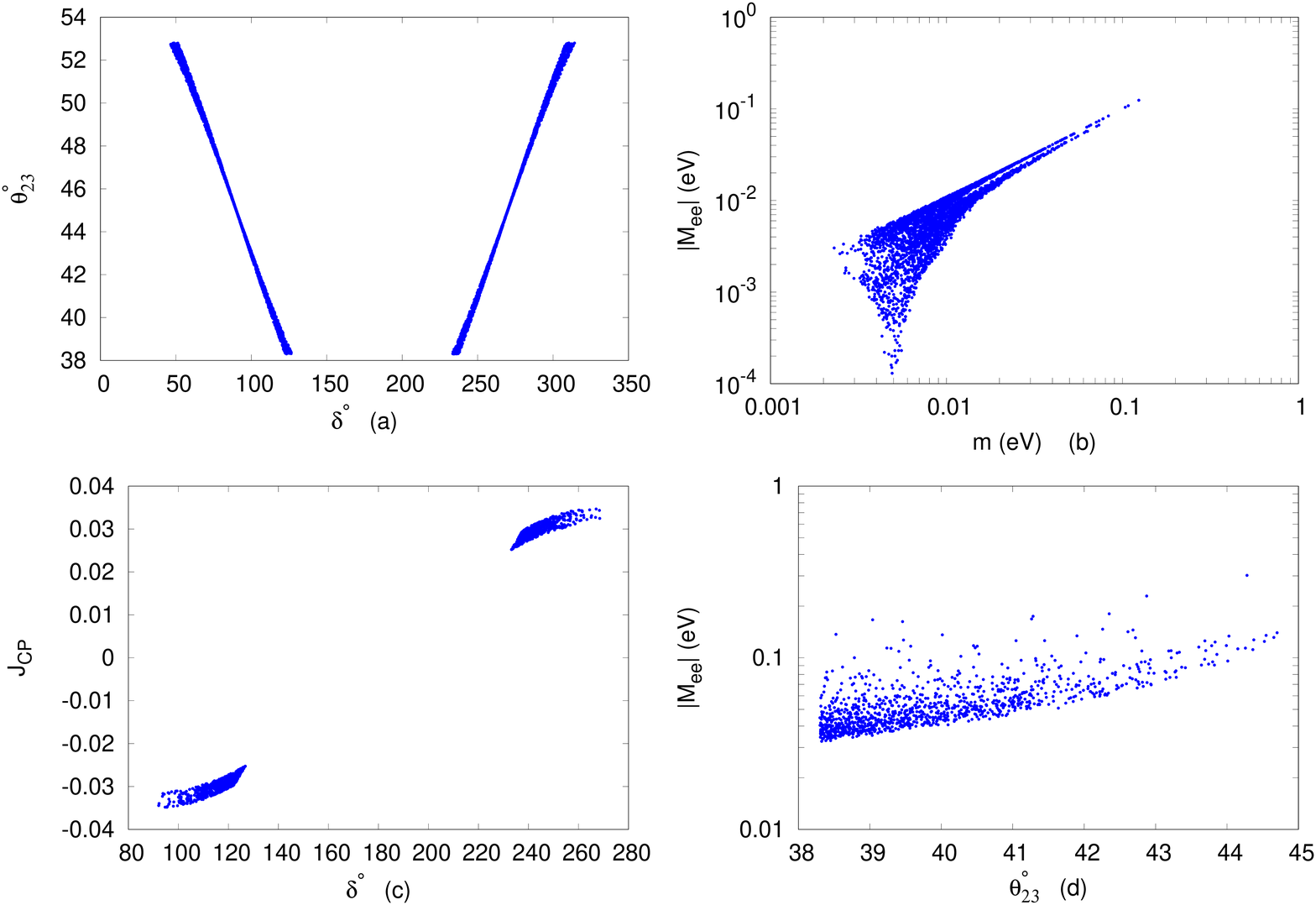}
\caption{Correlation plots for patterns $M^{\textrm{I}}_{\text{TM}_1}$ (a), $M^{\textrm{II}}_{\text{TM}_1}$ (b) and $M^{\textrm{IV}}_{\text{TM}_1}$ (c,d) with NO.}
\label{fig4}
\end{figure*} 
The main results of the numerical analysis are:
\begin{enumerate}[i]
\item All six patterns of one texture zero in the neutrino mass matrix with TM$_1$ mixing are consistent with the present neutrino oscillation data.
\item The pattern $M_{\text{TM}_1}^{\textrm{I}}$ is consistent with normal mass ordering only.
\item All the allowed patterns except for $M_{\text{TM}_1}^{\textrm{I}}$, allow a quasidegenerate mass spectrum.
\item In case of NO, vanishing values of the parameter $|M_{ee}|$ are allowed for patterns $M_{\text{TM}_1}^{\textrm{I}}$, $M_{\text{TM}_1}^{\textrm{II}}$ and $M_{\text{TM}_1}^{\textrm{III}}$. For the remaining patterns $|M_{ee}|$ is found to be bounded from below.
\item The smallest neutrino mass can have vanishing values for patterns $M_{\text{TM}_1}^{\textrm{II}}$ and $M_{\text{TM}_1}^{\textrm{III}}$ with IO.  
\item The parameter $J_{CP}$ cannot vanish for any of the allowed patterns implying that these patterns are necessarily CP violating.
\item The atmospheric neutrino mixing angle $\theta_{23}$ remains below (above) 45$^\circ$ for pattern $M^{\textrm{IV}}_{\textrm{TM}_1}$ ($M^{\textrm{VI}}_{\textrm{TM}_1}$) with NO.
\end{enumerate}
\begin{figure*}[t]
\centering 
\includegraphics[scale=0.3]{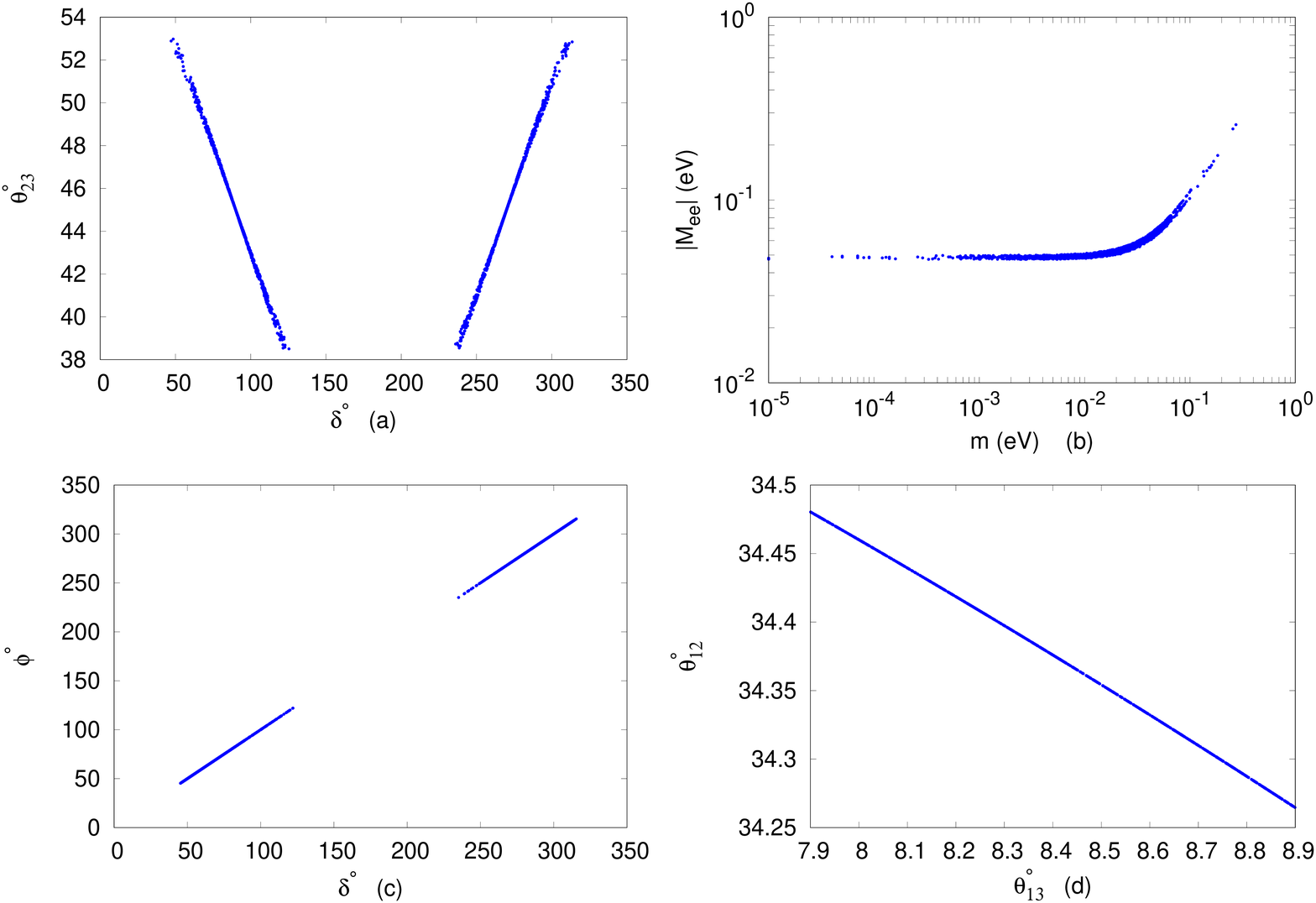}
\caption{Correlation plots for patterns $M^{\textrm{II}}_{\text{TM}_1}$ (a,b), $M^{\textrm{IV}}_{\text{TM}_1}$ (c) and $M^{\textrm{V}}_{\text{TM}_1}$ (d) with IO.}
\label{fig5}
\end{figure*} 
Numerical results for the presently unknown neutrino parameters have been summarized in Table \ref{tab5}. The allowed ranges (at 3$\sigma$ CL) of parameters $\theta_{12}$, $\theta$ and $J_{CP}$ are ($34.26^{\circ}$ - $34.48^{\circ}$), ($13.77^{\circ}$ - $15.55^{\circ}$) and $\pm$($0.022$ - $0.035$), respectively, for all the allowed patterns. Some of the correlation plots are given in Figs. \ref{fig4} and \ref{fig5}. Fig. \ref{fig4}(a) depicts the correlation plot between Dirac phase $\delta$ and mixing angle $\theta_{23}$ for patten $M^{\textrm{II}}_{\text{TM}_1}$ with NO. The CP violating Dirac phase $\delta$ is restricted to two regions around $90^{\circ}$ and $270^{\circ}$. This result holds for all the allowed patterns and is independent of the mass ordering. In the numerical analysis we have varied the Dirac phase $\delta$ within its full possible range of $0^\circ$ - $360^\circ$. Recent long baseline neutrino oscillation experiments such as MINOS and T2K  \cite{lbl} have shown a preference for the CP violating phase $\delta$ to be around 270$^\circ$. Particularly, recent global analysis in Ref. \cite{data} rules out $\delta$ from $32^\circ$ to $141^\circ$ at 3$\sigma$ CL for inverted mass ordering. If we take into account the limits on $\delta$ as given in Ref. \cite{data}, the region of $\delta$ around $90^\circ$ is ruled out and only the second region around 270$^\circ$, remains compatible with IO. Fig. \ref{fig4}(c) shows the correlation plot between $\delta$ and $J_{CP}$ for pattern $M^{\textrm{IV}}_{\text{TM}_1}$ with NO. It is clear that a vanishing $J_{CP}$ is not allowed for this pattern, in fact, all the patterns with one texture zero and TM$_1$ mixing predict a non-zero $J_{CP}$ which implies that these patterns are necessarily CP violating. This is because these patterns do not allow values $0^{\circ}$ and $180^{\circ}$ for the Dirac phase $\delta$ and since all the mixing angles are non-zero, the CP invariant $J_{CP}$ cannot vanish.\\ 
Phases $\phi$ and $\delta$ are found to have almost identical values [Fig. \ref{fig5}(c)] which is similar to the TM$_2$ case. 
The correlation between mixing angles $\theta_{13}$-$\theta_{12}$ is shown in Fig. \ref{fig5}(d). In contrast to the TM$_2$ case, here, the value of $\theta_{12}$ decreases with increasing $\theta$. This is a generic feature of TM$_1$ mixing arising from Eq. (\ref{eqth122}). This brings $\theta_{12}$ near its best fit experimental value. Thus TM$_1$  mixing is phenomenologically more appealing than TM$_2$ mixing.
\begin{table*}[t]
\centering
\begin{tabular}{|c|c|c|c|c|c|}
 \hline 
Pattern     & Mass Ordering & $|M_{ee}|$ (eV)    & $m$ (eV)           & $\Sigma m_i$ (eV) & $\delta^{\circ}$            \\
 \hline
I& NO       & $0.0$            & $0.0022$ - $0.0071$& $0.06$ - $0.07$  & $46$ - $127$ $\oplus$ $233$ - $314$ \\
             \hline 
II& NO     & $0.0$ - $0.124$  & $0.0017$ - $0.124$ & $0.06$ - $0.382$ & $46$ - $127$ $\oplus$ $233$ - $314$ \\
            & IO       & $0.047$ - $0.258$& $0.0$ - $0.273$    & $0.097$ - $0.83$ & $46$ - $127$ $\oplus$ $233$ - $314$ \\
 \hline
III& NO    & $0.0$ - $0.124$  & $0.0017$ - $0.124$ & $0.06$ - $0.382$ & $46$ - $127$ $\oplus$ $233$ - $314$ \\ 
            & IO       & $0.047$ - $0.258$& $0.0$ - $0.273$    & $0.097$ - $0.83$ & $46$ - $127$ $\oplus$ $233$ - $314$ \\
 \hline  
IV& NO  & $0.03$ - $0.3$ & $0.033$ - $0.3$    & $0.13$ - $0.91$  & $90$ - $127$ $\oplus$ $233$ - $270$ \\
            & IO       & $0.017$ - $0.24$ & $0.0037$ - $0.28$  & $0.1$ - $0.85$   & $45$ - $122$ $\oplus$ $235$ - $313$ \\
 \hline
V& NO & $0.14$ - $0.311$ & $0.15$ - $0.331$  & $0.46$ - $1$      & $48$ - $125$ $\oplus$ $235$ - $313$\\
            & IO       & $0.018$ - $0.24$ & $0.016$ - $0.26$   & $0.12$ - $0.78$  & $45$ - $125$ $\oplus$ $235$ - $314$ \\
 \hline
VI& NO& $0.029$ - $0.3$ & $0.031$ - $0.3$    & $0.12$ - $0.9$  & $45$ - $90$ $\oplus$ $270$ - $314$  \\ 
            & IO       & $0.017$ - $0.24$ & $0.0037$ - $0.28$  & $0.1$ - $0.85$   & $52$ - $125$ $\oplus$ $235$ - $314$ \\
 \hline
\end{tabular}
\caption{Numerical predictions (at 3$\sigma$ CL) for patterns having one texture zero in $M_\nu$ with TM$_1$ mixing.}\label{tab5}
\end{table*}
\section{Summary}
We studied the implications of having one texture zero in the neutrino mass matrix along with TM$_1$/TM$_2$ mixing. Considering neutrinos to be Majorana fermions, there are six possible patterns of one texture zero in the neutrino mass matrix. All the six patterns are found to be phenomenologically allowed when combined with TM$_1$/TM$_2$ mixing. The presence of a texture zero in the neutrino mass matrix leads to relations between neutrino masses and mixing matrix elements whereas TM$_1$/TM$_2$ mixing implies relations between mixing angles. Thus, the combination of one texture zero patterns with TM$_1$/TM$_2$ mixing leads to very predictive neutrino mass matrices. For the pattern where the texture zero is at (1,1) position in the neutrino mass matrix, only normal mass ordering is experimentally allowed. Since TM$_2$ mixing predicts values of $\theta_{12}$ away from its best fit value, TM$_1$ mixing is phenomenologically more appealing. We have obtained predictions for the unknown parameters such as the effective Majorana neutrino mass, the Dirac CP violating phase and the neutrino mass scale. The Dirac phase $\delta$ has been found to be strongly correlated with the phase parameter $\phi$ for both TM$_1$ as well as TM$_2$ mixing. We have also constructed neutrino mass models which lead to patterns of one texture zero with TM$_2$ mixing. To realize these patterns we have employed the $A_4$ symmetry within the framework of type-I+II seesaw mechanism.

\acknowledgements{
R. R. G. acknowledges the financial support provided by the Council of Scientific and Industrial Research (CSIR), Government of India, Grant No. 13(8949-A)/2017-Pool. Part of this work was supported by the Department of Science and Technology, Government of India, Grant No. SB/FTP/PS-128/2013. I thank Sanjeev Kumar and Desh Raj for carefully reading the manuscript.}

\appendix*
\section{Group $A_4$}
$A_4$ has four inequivalent irreducible representations (IRs) which are three singlets \textbf{1}, $\textbf{1}^\prime$, and $\textbf{1}^{\prime\prime}$, and one triplet \textbf{3}. The group $A_4$ is generated by two generators $S$ and $T$ such that 
\begin{equation}
S^2 = T^3 = (S T)^3 = 1.
\end{equation}
The one dimensional unitary IRs are
\begin{equation}
\textbf{1} \  S = 1 \ \ T = 1, \ \
\textbf{1}^{\prime} \  S = 1 \ \ T = \omega, 
\textbf{1}^{\prime\prime} \  S = 1 \ \ T = \omega^2.
\end{equation}
The three dimensional unitary IR in the $S$ diagonal basis is 
 \begin{equation}
 S = \left(
\begin{array}{ccc}
1 & 0 & 0 \\ 0 & -1 & 0 \\ 0& 0 & -1
\end{array}
\right), \ \ T = \left(
\begin{array}{ccc}
0 & 1 & 0 \\ 0 & 0 & 1 \\ 1& 0 & 0
\end{array}
\right).
\end{equation}
The multiplication rules of the IRs are as follows
\begin{equation}
\textbf{1}^\prime \otimes \textbf{1}^\prime = \textbf{1}^{\prime \prime}, \ \textbf{1}^{\prime \prime} \otimes \textbf{1}^{\prime \prime} = \textbf{1}^{\prime}, \ \textbf{1}^{\prime} \otimes \textbf{1}^{\prime \prime} = \textbf{1}.
\end{equation}
The product of two $\textbf{3}$'s gives 
\begin{equation}
\textbf{3} \otimes \textbf{3} = \textbf{1} \oplus \textbf{1}^\prime \oplus \textbf{1}^{\prime \prime} \oplus \textbf{3}_s \oplus \textbf{3}_a,
\end{equation}
where $s$, $a$ denote the symmetric, anti-symmetric products, respectively. 
Let $(x_1, x_2, x_3)$ and $(y_1, y_2, y_3)$ denote 
the basis vectors of two $\textbf{3}$'s. IRs obtained from 
their products are
\begin{align}
(\textbf{3}\otimes\textbf{3})_{\textbf{1}} & = x_1 y_1 + x_2 y_2 + x_3 y_3    \\
(\textbf{3}\otimes\textbf{3})_{\textbf{1}^\prime} & = x_1 y_1 + \omega^2 x_2 y_2 + \omega x_3 y_3   \\
(\textbf{3}\otimes\textbf{3})_{\textbf{1}^{\prime \prime}} & = x_1 y_1 + \omega x_2 y_2 + \omega^2 x_3 y_3   \\
(\textbf{3}\otimes\textbf{3})_{\textbf{3}_s} & = (x_2 y_3 + x_3 y_2, x_3 y_1 + x_1 y_3, x_1 y_2 + x_2 y_1)   \\
(\textbf{3}\otimes\textbf{3})_{\textbf{3}_a} & = (x_2 y_3 - x_3 y_2, x_3 y_1 - x_1 y_3, x_1 y_2 - x_2 y_1).
\end{align}

\end{document}